# Uncompensated-spins Induced Weak Ferromagnetism in $Ca_3Mn_2O_7$: Magneto-conductive and dual Magneto-capacitive Effects

Pooja Sahlot and A.M. Awasthi*

*UGC-DAE Consortium for Scientific Research, University Campus, Khandwa Road, Indore- 452 001, India*

*amawasthi@csr.res.in

**Abstract**

Temperature dependent magnetization study on single phase orthorhombic Ruddlesden-Popper manganite $Ca_3Mn_2O_7$ evidences antiferromagnetic (AFM) ordering below 123K. Field-dependent magnetization $M(H)$ depicts off-centered hysteretic loops below ~110K— confirming the existence of both weak ferromagnetism (WFM) and exchange bias, whose development upon cooling is examined in detail. WFM is attributed to the formation of uncompensated-spin-clusters within the AFM-matrix, stabilized by the high-anisotropy in the manganite, favoring Dzyaloshinskii-Moriya (D-M) interaction. Temperature dependence of an evaluated average canting angle from the $M(H)$ loops provides a quantitative measure of the WFM-evolution. Dielectric measurements exhibit magneto-conductive effects concurrent to the WFM-onset. Temperature dependence of the low-frequency/dc-limit of Jonscher-fits to the conductivity isotherms confirms the presence of relatively more-conductive clusters, embedded in the insulating-bulk. Nyquist impedance-analysis of complex impedance reveals two relaxations, yielding dual magnetic-field effects; the lumped circuit parameters feature anomalies across the magnetic phase changes.

## 1. Introduction

Lately, octahedral reorientations in perovskites have been a topic of continuous interest, because of the strong coupling with magnetic properties, owing to the structure dependence of exchange interactions between the magnetic ions [1]. Existence of exchange bias is one of the interesting magnetic properties with various practical applications [2]. Antiferromagnets (AFMs) possess large anisotropy and ferromagnets (FMs) are known to possess large exchange parameter. Interfacial coupling between these configurations yields unidirectional feature to the FM part, by pinning of the FM spins; resulting in shift of the $M(H)$ hysteresis loop, referred to as exchange bias (EB) [3]. This pinning is effected by the uncompensated AFM moments at the interface. After the first report discussing field-cooled (FC)-EB in bulk manganites, possessing dispersed FM regions in the AFM base, many bulk systems have been reported showing EB [4, 5]. Apart from the FC-EB, Saha et. al. [6] discussed shift in the hysteresis loop for the zero-field cooled systems i.e., (ZFC)-EB. Further, magneto-electric coupling driven exchange bias has been established in the bulk compensated-AFM $BaMnF_4$ by Zhou et. al. [7]. Dong et. al. [8] reported a model that G-AFM/FM interfaces, formed by oxygen octahedron tilting in oxide hetero-structures, result in spin-pinning and thus the EB. This is suggested to be caused by the D-M interaction and ferroelectric polarization in multiferroics.

On heading to perovskite-related layered-structures, namely the Ruddlesden-Popper (R-P) compounds, stronger intra-layer correlations are expected, as there is reduction in dimensionality of the network of corner-sharing octahedra. R-P compounds with the general formula $A^{2+}_{n+1}B^{4+}_n O^{2-}_{3n+1}$ are structured as blocks of '$n$' (integer) number of $ABO_3$ perovskite-like octahedron layers, with intermediary AO layers, whereas the general perovskites are pictured for $n =\infty$ [9, 10]. In some $n =2$ R-P compounds, combination of octahedral reorientations has been predicted to introduce mixed magnetic states [11]. Also, FM-clustering in AFM insulators has been demonstrated in slightly La-doped $Sr_4Mn_3O_{10}$ $n =2$ R-P system [12]. $Ca_3Mn_2O_7$ undergoes phase transition from high-temperature tetragonal crystal structure into low-temperature orthorhombic structure, with the introduction of two coupled lattice modes; oxygen octahedral tilting and oxygen octahedron rotation [11, 13]. Nicole et. al. [13] determined from first principles calculations for $Ca_3Mn_2O_7$ system, to possess G-type AFM ground state, with specified direction of spins, associated with the high crystalline anisotropy [11, 13]. Below the AFM ordering, additional spin-orbit interactions, because of the octahedron tilt, were said to give rise to a net spin moment perpendicular to the staggered magnetic vector, estimated at $0.18\mu_B$ per unit cell [11]. Later, a report on the neutron study of $Ca_3Mn_2O_7$ established these predictions [14]. Antiferromagnetic transition with Nèel temperature ~115K was reported in the compound. Low temperature Neutron diffraction data evidenced weak ferromagnetism (WFM) in the G-type AFM ordering. Transition to the AFM state with $T_N$ =134K in $Ca_3Mn_2O_7$, along with the emergence of WFM below 100K, has been shown in its magnetization [15]. Application of Dzyaloshinskii-Moriya criteria [13, 16, 17] shows that the net moment (WFM) results from the oxygen octahedron-tilt distortions. $Ca_3Mn_2O_7$ shows exchange bias upon cooling, which evidences the emergence of uncompensated spin-clusters (WFM) in the AFM-base matrix, and thereby the D-M interaction [18]. In the present script, temperature, and field-dependent magnetization has been investigated for $Ca_3Mn_2O_7$. Detailed study of EB and WFM characters with temperature has been done. Frequency & temperature dependences of measured ac-conductivity have been analyzed whose parametric characterization corroborates the evolution of magnetic configuration. Complex impedance analysis has been carried out revealing different magnetic field effects on the lumped-circuit elements, representing the low- and high- frequency relaxations.

## 2. Experimental Details

Practicing conventional solid state synthesis technique, single phase polycrystalline $Ca_3Mn_2O_7$ was synthesized at 1300 °C using $CaCO_3$ and $MnCO_3$ precursors and structurally characterized by powder XRD using Bruker D8 advance diffractometer with Cu-$K_\alpha$ X-ray radiation ($\lambda$ =1.5405 Å). By room temperature XRD, the prepared specimen was characterized having orthorhombic structure, with lattice parameters $a$ =19.40(8) Å, $b$ =5.24(3) Å, and $c$ =5.25(2) Å, as reported elsewhere [18]. The iodometric titration performed yielded the close stoichiometry of the specimen as $Ca_3Mn_2O_{6.977}$. Using a Quantum Design SQUID-VSM, zero-field cooled (ZFC) and field cooled (FC) $M(T)$ and low temperature $M(H)$ were measured. AC-conductivity and complex impedance measurements were performed using Novocontrol Alpha-A Broadband Impedance Analyzer and Oxford Nanosystem's Integra 9T magnet-cryostat.

## 3. Results and Discussion

### 3.1. Magnetic properties

Figure.1 (left $y$-axis) shows zero field cooled (ZFC) dc-magnetic susceptibility ($\chi$ =$M/H$) vs. temperature ($T$) measurement under 100 Oe field for $Ca_3Mn_2O_7$. Mn-spins' short-range ordering induced broad hump is observed around 155K, consistent with the published reports [19]. Well-built correlations induce the AFM ordering below 123K. At temperatures well above the AFM ordering, the expected Curie-Weiss behavior $\chi$ =$C/(T-\vartheta_{C-W})$, is seen in the inverse susceptibility $1/\chi$ vs. temperature plot (fig.1 inset). The fitting yields -ve $T$-axis intercept $\vartheta_{C-W}$ = -889K, which further confirms the mentioned AFM-ordering, and the Curie constant is found to be $C$ =0.0217 K-emu/(gm-Oe). The effective magnetic moment $\mu$ of $Mn^{+4}$ ions is involved in the Curie constant as $C$ =$N_A\mu^2/(3A_{Mn}k_B)$, where $N_A$ is the Avogadro number, $k_B$ is the Boltzmann constant, and $A_{Mn}$ is the atomic mass of the magnetic ion. Hence obtained value of $\mu$ =3.09$\mu_B$ is fairly close to the theoretically anticipated [20] moment 3.84$\mu_B$ for the $Mn^{+4}$ ions. Temperature dependence of dc magnetic susceptibility $\chi$ shows a change (maximum) in slope at ~110K, below the AFM ordering at $T_N$ =123K, with a clear upturn in susceptibility below 60K. The extra slope-extremum ($dM/dT$ in fig.1, right $y$-axis) signals the emergence of a WFM character in the AFM-matrix state.

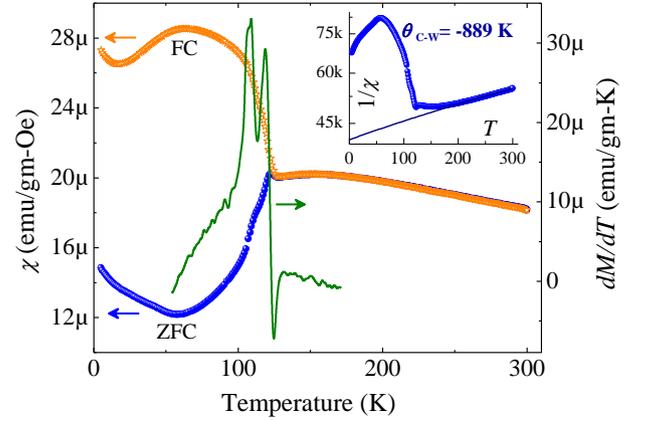

Fig.1 ZFC/FC-DC magnetic susceptibility $\chi(T)$ (left $y$-axis) under 100 Oe field and ZFC-$dM/dT$ (right $y$-axis), showing an additional slope-anomaly at ~110K, below that due to the AFM-$T_N$ at 123K; Bottom/right inset: ZFC-$1/\chi(T)$ under 100 Oe field, with its high-$T$ linear (Curie- Weiss Fit) above 225K.

Fig.1 shows FC $\chi$-$T$ curve (taken from reference [18]), along with ZFC $\chi$-$T$ under 100 Oe field. Considerable splitting of the low-field ZFC/FC curves seen below the ~125K indicates the presence of frustration in the magnetic state. This bifurcation further enhances with decrease in temperature, because of the D-M interaction induced canted/uncompensated spins. As the temperature is increased, thermal fluctuations support rotation of some of the spins along the direction of the applied magnetic field, leading to the reduction in split with temperature in $\chi(T)$ under 100 Oe field. Upturn seen below ~15K under 100 Oe FC is contribution from the paramagnetic/unpaired spins at low temperatures.

To probe the magnetic character (below AFM transition) of the system in detail, the field dependent magnetization isotherms taken at low temperatures are shown in fig.2(a), (b). Observed maxima in slope of $M(H)$ isotherms ($dM/dH$, descending curve, fig.2(c), (d)) at higher fields ($H_{SR}$) suggest spin reorientations near ~50 kOe. $H_{SR}$ increases below $T_N$, reaching a maximum near 75K (fig.2(e)). Barrier-enhancement of field-driven spin reorientation with departure from $T_N$ has been observed in other systems [21]. Decreasing $H_{SR}$ with further cooling below 75K explains the dominance of magneto-crystalline anisotropy over thermal fluctuations at low temperatures, also observed in FC-$\chi(T)$ (fig.1). On lowering temperature below 110K, the system shows clear hysteretic nature, along with the exchange-bias (EB) effect, as evident from the shift of $M$-$H$ isotherms towards the negative field-direction from $H$ =0 center, as reported previously [18].

For a comprehensive analysis presented here, $M(H)$ isotherms were further recorded at yet lower temperatures; albeit, because of their primary AFM nature, hysteretic $M(H)$ loops escape magnetization saturation at high fields. Nonetheless, overlap in descending and ascending curves of the isotherms at high-fields (fig.2(a)) indicates the 'effective saturation' of the WFM-subsystem [22]. This eliminates the possibility of minor-loop effects under low maximum fields, as unable to reverse the FM moments. Also, here the maximum applied field of 70 kOe is much higher than the observed coercivity ($\leq O$(800 Oe), fig.2(b)) of the identified WFM subsystem. Concurrent appearance of the exchange-bias with the onset temperature of WFM below ~105K suggests their inter-dependence. In the bulk $Ca_3Mn_2O_7$, nucleation of clusters of uncompensated-spins in the AFM-matrix state, giving WFM below 110K is claimed [18]. With further cooling, octahedral correlations enhance the growth of localized WFM clusters.

Large bifurcation of FC and ZFC $M$-$T$ curves, hysteresis behaviour in $M(H)$ isotherms, and the small remnant magnetization observed below the ordering temperature, are all consistent with the formation of magnetic clusters of uncompensated moments. EB in $Ca_3Mn_2O_7$ is a bulk effect; attributed to the pinning of WFM, interfacially coupled to AFM, caused by the D-M interaction, supported by theoretical work of Dong et. al. [8]. Hence, the manifestation of exchange bias in $Ca_3Mn_2O_7$ is a signature of the D-MI-induced uncompensated-spins clusters, and we further analyze the same as follows.

The coercive field $H_c$ corresponding to the WFM character of the system is obtained from the $M$-$H$ loops shown in fig.2(b) at different temperatures. $H_c$ reduces with the increase in thermal energy, suppressed by the fluctuations induced in the anisotropy-driven spins. With the formation of WFM clusters below ~110K, $H_c(T)$ grows exponentially on lowing temperature, as shown in fig.2(f).

The exchange bias in the system is characterized by the horizontal shift $H_{hs}$ in $M(H)$ isotherms;

$$H_{hs} = \frac{|H_{c+} + H_{c-}|}{2},$$

$H_{hs}$ as defined is obtained from $M(H)$ isotherms (Fig.2(b)) for different temperatures. $H_{hs}$ is found to increase almost linearly initially below 110K, which is explained by taking into account thermal consolidation of spins' canting upon cooling. [23]. At low temperatures, $H_{hs}$ tends to saturate; accordingly $H_{hs}(T)$ fit in fig.2.(g) exhibits the power-law dependence as ~$(T-108)^{0.8}$; the fitted 108K is close to the WFM-emergence temperature.

The behaviour of saturation magnetization with temperature obtained for $M(H)$ isotherms (fig.2 (a)) is also explored. For bulk ferromagnets, well known $T^{3/2}$ law at low temperatures is given by Bloch [24]. Although, for other anisotropic magnetic systems like nanoparticles, modified Bloch's law with deviation in the value of Bloch's exponent $\alpha_B$ has been suggested [25] as

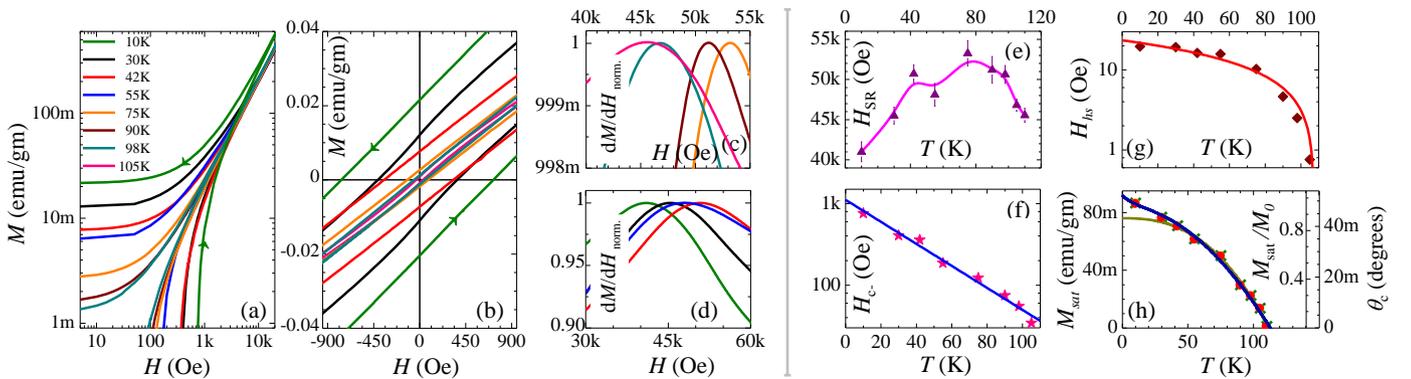

Fig.2 (a) Log-log plots of the $M(H)$ measurements showing overlap of ascending/descending branches at various temperatures. (b) Zoomed-in view of $M(H)$ at different temperatures, showing the steady development of WFM hysteresis. (c) & (d) Normalized $dM/dH$ vs. $H$ plots, showing peaks corresponding to the spin reorientation near 50 kOe field for 10K, 30K, 42K, and 55K and for 75K, 90K, 105K, and 110K, respectively. (e) Temperature dependence of the spin-reorientation field ($H_{SR}$), at $dM/dH$-maxima in (c), (d). (f) Temperature dependence of the -ve coercive field $|H_{c-}(T)|$ on logarithmic $y$-scale. (g) Horizontal shift ($H_{hs}$ versus $T$) of the loop-center. (h) Temperature dependence of $M_{sat}$ (fit in yellow) on left pane, with temperature dependence of $M_{sat}/M_0$ (fit in blue) and $\theta_c$ (right pane) as a function of temperature.

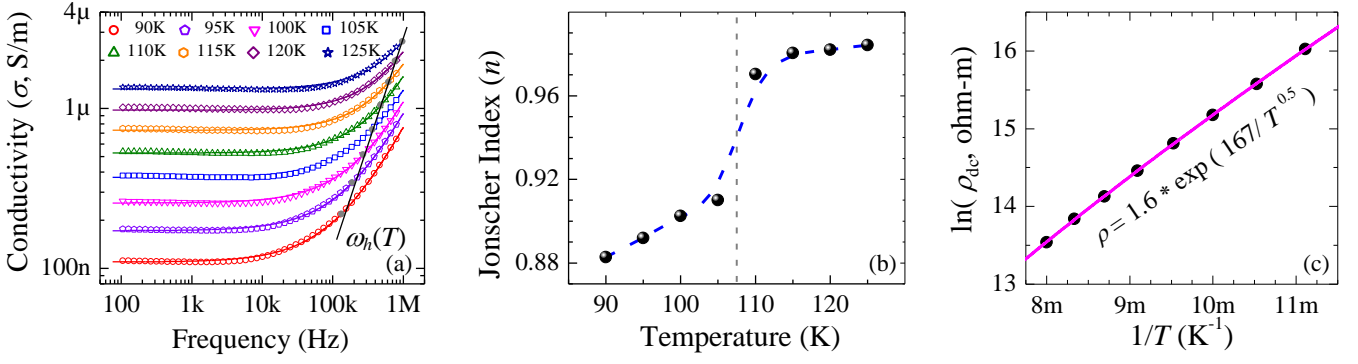

Fig.3.(a) Log-log plot of conductivity ($\sigma$) vs. frequency ($f$) for temperatures 90K, 95K, 100K, 105K, 110K, 115K, 120K, and 125K.
(b) Temperature dependence of the Jonscher power-law exponent $n(T)$ (left-bottom axes) and dc-conductivity $\sigma_{dc}(1/T)$ (right-top axes).
(c) Theoretical fit [35] on the dc-resistivity $\rho_{dc}(T) = 1/\sigma_{dc}(T)$.

$$M_{sat} = M_s(0)\left(1-(T/T_0)^{\alpha_B}\right)$$

Here, for the mentioned system, best fit for temperature variation of +ve saturation magnetization $M_{sat}(T)$ for the shown $M(H)$ isotherms is shown in fig.2 (h): left axis, with 0K saturation magnetization $M_s(0)$ =0.079 emu/gm, for ordering temperature $T_0$ =111.6K, with $\alpha_B$ =2.5. Anisotropy in the magnetic structure of the system explains the deviation in the exponent-value (from 3/2), as also reported for other bulk and nano systems [26, 27]. The finite-size effects of WFM clusters and spin-canting effect are deemed to play a crucial role in this departure of the WFM magnetization-saturation behaviour from the original equation. Surface contribution to the core magnetization is prominently observed at low temperatures. This has been explained via addition of a surface term, corresponding to the modulated spins across the WFM-surface [28];

$$\frac{M_{sat}(T)}{M_s(0)} = \left(1-(T/T_0)^{\alpha_B}\right) + S_0 \text{Exp}\left(\frac{-T}{T_f}\right)$$

Surface contributory parameter $S_0$ is obtained to be 0.09 and freezing temperature of the surface spins is ~8.4K. Also, the averaged spin canting ($\theta_c$) in the system is examined with decrease in temperature, obtained from the ratio of $M_{sat}$ at corresponding temperature to the maximum moment of $Mn^{+4}$ ions in the system, latter evaluated using the C-W fit. The obtained maximum canting angle is in agreement with an earlier reported value (~0.02°) for the system [19]. Figure 2 (h) right-pane shows hence obtained values of $\theta_c(T)$, depicting increase in $\theta_c$ with decrease in temperature, which replicates the behaviour of $M_{sat}(T)$. The fitted value of $T_0$ =111.6K for $\theta_c(T)$ agrees fairly with the WFM-character induced in the system below ~110K, as observed in $M(T)$ data.

### 3.2. AC and DC conductivity

Electrical conductivity in the system is described as the combination of dc- and ac- components [29]. Flat/frequency-independent regime corresponds to the dc-conductivity ($\sigma_{dc}$) and frequency-dependent part is ac-conductivity, which follows the Jonscher power law: $\sigma_{ac} = A\omega^n$ [30]. Here, $\omega = 2\pi f$ is alternating electric field's angular frequency, $A$ is pre-exponential factor, and $s$ is the power-exponent whose behaviour vs. temperature describes the conduction mechanism in the system. Here, conductivity in the system has been explained using the "universal law of dielectrics", in appreciable range of temperature and frequency:

$$\sigma(\omega) = \sigma_{dc} + A\omega^n = \sigma_{dc}\left(1+(\omega/\omega_h)^n\right)$$

With increase in temperature, the crossover frequency $\omega_h$ from dc- to ac-conduction ($\sigma(\omega_h) = 2\sigma_{dc}$), shifts to higher values. Jonscher fits for $\sigma(f)$ plot at different temperatures shown in fig.3(a) give temperature dependent exponent $n(T)$ and $\sigma_{dc}(T)$. Exponent $n(T)$ shown in fig.3(b) encounters down-step, concurrent with the WFM-onset at 110K, with enhanced slope at lower $T$'s. Evolution of the magnetic state in $Ca_3Mn_2O_7$, associated with lattice distortion below 110K, affects the energy distribution and hopping distance of charge carriers [31, 32]. The higher values of Jonscher exponent observed here, registers significant role of multiple hopping [33, 34]. At appreciably low frequencies, Jonscher power-law fitted $\sigma_{dc}$ follows the resistivity ($\rho_{dc} = 1/\sigma_{dc}$) behaviour of the following form, shown in fig.3(c);

$$\rho_{dc}(T) = \rho_0 \text{Exp}\left(\sqrt{T_0/T}\right)$$

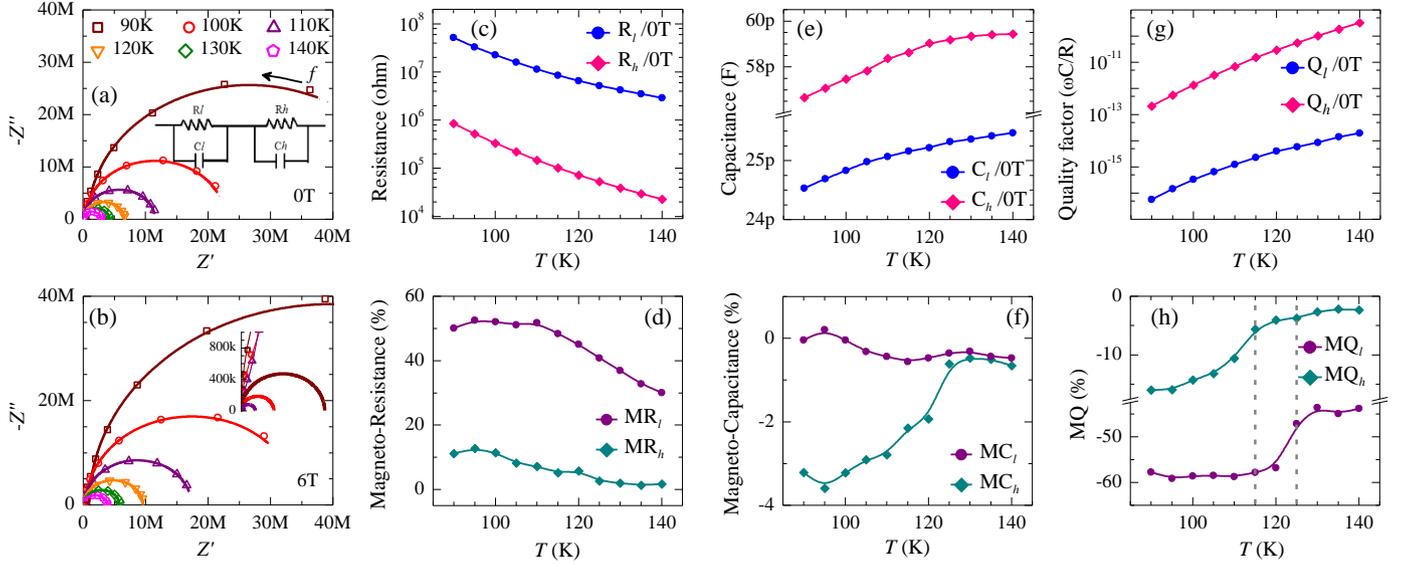

Figure 4. (a),(b) Nyquist plots of $-Z''(\omega)$ vs. $Z'(\omega)$ with the fitted curves. Inset in (b): High frequency Nyquist fit. (c), (d) Lumped-circuit resistance and MR of low- and high- frequency relaxations respectively. (e), (f) Lumped-circuit capacitance and MC of low- and high-frequency relaxations respectively. (g), (h) Lumped-circuit quality factor ($Q$) and MQ of low- and high-frequency relaxations respectively.

This is well explained by the clustering theory reflecting electrical conduction occurring via hopping of the charge carriers from 'metallic' to insulating regions., given by Sheng et. al. [35]. In $Ca_3Mn_2O_7$, WFM-metallic nano-clusters are embedded in the AFM-insulating matrix, which here evidences with $\ln\rho_{dc} \sim T^{-1/2}$ dependence.

### 3.3. Impedance Spectroscopy

To distinguish relaxations for pinned-dipoles (low-frequencies) and free-dipoles (high-frequencies) [36,37], Nyquist modeling is performed with a lumped equivalent circuit shown in fig.4(a)-inset. It consists of a series combination of two resistance & capacitance parallel configurations. Fig.4(b) depicts the Nyquist-plot of impedance measured under 6T magnetic field. The two (depressed) semicircles reflect the non-Debyean nature of both the dipolar relaxations. Temperature variation of the lumped resistance and capacitance circuit-elements, evaluated from fitting the semi-circle at lower frequencies ($R_l$, $C_l$) and at high-frequencies ($R_h$, $C_h$), is shown in fig.4 (c) and (e). Magneto-resistance ($MR_l =\{R_l(H)-R_l(0)\}/R_l(0)$, $MR_h$) and similarly-defined magneto-capacitance ($MC_l$, $MC_h$) evaluated versus temperature are shown in fig.4 (d) and (f). Here, high-frequency semicircle corresponds to the dipole dynamics in the wake of WFM-nanophase interior field, whereas the low-frequency semicircle signifies dipolar relaxations in the bulk AFM-matrix [38]. $MR_l$ characterizing the bulk-AFM region becomes nearly temperature independent below 110K, whereas $MC_l$ tends to become less negative. Hence the low-frequency dipolar relaxations are not accompanied by the hopping of charge carriers. $MR_h$ reflecting the WFM-character is considerably less but does not saturate upon cooling, even as $MC_h$ becomes more negative. Hence the high-frequency relaxations accompany hopping of the charge carries.

Lumped RC-circuits considered here represent 'overdamped' limits of the dispersive dipolar-relaxations. Nevertheless, we may define the equivalent lumped quality-factors $Q$ for them as follows. In RC-circuit terms, resistance $R$ measures electrical-leakage (losses $\sim I^2R$) and susceptance $C\omega$ measures the charge-storage (per cycle). Therefore, the combination $C\omega_{ch}/R$ qualifies as a measure of $Q$ for the lumped RC-configuration; $\omega_{ch}$ being certain unique/characteristic frequency, which we pick here as that corresponding to the semicircles'-top. $Q = C\omega_{ch}/R$ evaluated for the low- and high-frequency lumped RC-circuits and their corresponding magneto-quality $MQ =\{Q(H)-Q(0)\}/Q(0)$ are shown in fig.4 (g) and (h), respectively. Consistent with the distinguished behaviors of $MC_l(T)$ and $MC_h(T)$, attributed to the bulk AFM-matrix and the WFM-nanophase respectively, $MQ_l(T)$ and $MQ_h(T)$ clearly register step-down anomalies at $\sim T_N$ and $\sim T_W$.

Finally, the anti-regression between the corresponding $R$ and $C$ lumped-circuit parameters observed here (vs. their variation against temperature), together with no regression between corresponding MR's and MC's both affirm that the magneto-capacitance here is unrelated to its magneto-resistance [39], evidencing existence of genuine magneto-dielectric coupling in the system.

The small oxygen vacancy ($Ca_3Mn_2O_{7-\delta}$, $\delta \approx 0.023$) determined in our specimen can nominally decrease the sample impedance by affecting its ac-conductivity, via activated *local* hole-hopping. Therefore, besides little influence on the dc-transport, our analysis carried out here and inferences drawn from the same are largely unaffected, as the oxygen vacancy is rather small.

## 4. Conclusions

Weak ferromagnetism (WFM) seen below $T_W$ =110K in Ruddlesden-Popper $Ca_3Mn_2O_7$ antiferromagnet ($T_N$ =123K) was studied and the temperature evolution of a discernible exchange bias was examined. WFM saturation-$M_{sat}$(WFM) follows the modified Bloch's law, with $T$-exponent $\alpha_B$ =2.5. AC electrical conductivity reveals the multiple-hopping mechanism, and the abrupt step-change in its Jonscher power-law exponent ($n$), concurrent with the WFM-emergence, signifies the magneto-conductive effect. Nyquist analysis of impedance affirms the dual nature of magneto-capacitance, attributed to the low- and high-frequency relaxations. Moreover, the system's AFM-transition and realization of WFM-nanophase within the bulk AFM-matrix are precisely benchmarked by step-anomalies, in the appropriately-defined (magneto) quality-factors of the lumped RC-circuits, representing the low- and high-frequency dipolar relaxations, respectively.

### Acknowledgments

We thank Mr. A. Jana and Ms. M. Tripathi for help with magnetization measurements, and appreciate Dr. R.J. Choudhary for their critical discussions. We extend our sincere thanks to Mr. Suresh Bhardwaj for help with the ac-conductivity measurements.

## References

[1] J.B. Goodenough, A. Wold, R.J. Arnott, and N. Menyuk, Physical Review **124** (2) (1961) 373.

[2] W.H. Meiklejohn and C.P. Bean, Physical Review **105** (3) (1956) 904.

[3] R.L. Stamps, Journal of Physics D: Applied Physics **33** (23) (2000) R247.

[4] D. Niebieskikwiat and M.B. Salamon, Physical Review B **72** (17) (2005) 174422.

[5] V. Markovich, I. Fita, A. Wisniewski, R. Puzniak, C. Martin, G. Jung, and G. Gorodetsky, Materials Chemistry and Physics **184** (2016) 49-56.

[6] J. Saha and R.H. Victora, Physical Review B **76** (10) (2007) 100405.

[7] S. Zhou, J. Wang, X. Chang, S. Wang, B. Qian, Z. Han, Q. Xu, J. Du, P. Wang, and S. Dong, Scientific Reports **5** (2015) 18392.

[8] S. Dong, K. Yamauchi, S. Yunoki, R. Yu, S. Liang, A. Moreo, J. M. Liu, S. Picozzi, and E. Dagotto, Physical Review Letters **103** (12) (2009) 127201.

[9] S.N. Ruddlesden and P. Popper, Acta Crystallographica **10** (8) (1957) 538-539.

[10] S.N. Ruddlesden and P. Popper, Acta Crystallographica **11** (1) (1958) 54-55.

[11] A.B. Harris, Physical Review B **84** (6) (2011) 064116.

[12] Y.K. Tang, X. Ma, Z.Q. Kou, Y. Sun, N.L. Di, Z.H. Cheng, and Q.A. Li, Physical Review B **72** (13) (2005) 132403.

[13] N.A. Benedek and C.J. Fennie, Physical Review Letters **106** (10) (2011) 107204.

[14] M.V. Lobanov, M. Greenblatt, N.C. El'ad, J.D. Jorgensen, D.V. Sheptyakov, B.H. Toby, C.E. Botez, and P.W. Stephens, Journal of Physics: Condensed Matter **16** (29) (2004) 5339.

[15] W.H. Jung, Journal of Materials Science Letters **19** (22) (2000) 2037-2038.

[16] I. Dzyaloshinskii, J. Phys. Chem. Solids **4** (1958) 241.

[17] T. Moriya, Phys. Rev. **120** (1) (1960) 91.

[18] P. Sahlot, A. Jana and A.M. Awasthi, AIP Conference Proceedings, **1942** (1) (2018) 130009.

[19] M.V. Lobanov, S. Li, and M. Greenblatt, Chemistry of Materials **15** (6) (2003) 1302-1308.

[20] A.I. Mihut, L.E. Spring, R.I. Bewley, S.J. Blundell, W. Hayes, T. Jestädt, B.W. Lovett, R. McDonald, F.L. Pratt, J. Singleton, and P.D. Battle, Journal of Physics: Cond. Matter **10** (45) (1998) L727.

[21] S.S. Samatham, A.K. Patel, and K.G. Suresh, arXiv preprint arXiv:1711.03263 (2017).

[22] J. Geshev, Journal of Physics: Condensed Matter **21** (7) (2009) 078001.

[23] T. Bora, P. Saravanan, and S. Ravi, Journal of superconductivity and novel magnetism, **26** (5) (2013) 1645-1648.

[24] F. Bloch, Z. Phys. (Zeitschrift für Physik) **61** (3-4) (1930) 206-219.

[25] K. Mandal, S. Mitra, and P.A. Kumar, Europhysics Letters **75**(no. 4), (2006) 618.

[26] C. Vázquez-Vázquez, M.A. López-Quintela, M.C. Buján-Núñez, and J. Rivas, Journal of Nanoparticle Research **13**(4) (2011) 1663-1676.

[27] B.K. Chatterjee, A. Dey, C.K. Ghosh, and K.K. Chattopadhyay, Journal of Magnetism and Magnetic Materials **367** (2014) 19-32.

[28] Vázquez-Vázquez, C., López-Quintela, M. A., Buján-Núñez, M. C., & Rivas, J. (2011). Journal of Nanoparticle Research, **13**(4), 1663-1676.

[29] A. Ghosh, Physical Review B **41**(3) (1990) 1479.

[30] A.K. Jonscher, Nature **267** (5613) (1977) 673.

[31] M. Pollak, Physical Review **133** (2A) (1964) A564-A579.

[32] M. Pollak, Philosophical Magazine **23** (183) (1971) 519-542.

[33] M. Pollak, Journal of Physics C: Solid State Physics **14** (21) (1981) 2977.

[34] S. Abboudy, K. Alfaramawi, and L. Abulnasr, Modern Physics Letter B **28** (01) (2014) 1450002.

[35] Sheng, P., Abeles, B., & Arie, Y., Physical Review Letters, **31**(1) (1973) 44.

[36] Jonscher, A.K., Nature, **253**(5494) (1975) 717.

[37] Kumar, N.K., Shahid, T. S., & Govindaraj, G., Physica B: Condensed Matter, **488** (2016) 99-107.

[38] Sahlot, P. and Awasthi, A.M., arXiv preprint arXiv:1812.01418 (2018).

[39] Catalan G., Appl. Phys. Lett. **88** (2006) 102902.